\documentclass[secnumarabic,amssymb, nobibnotes]{revtex4-2}

\usepackage{graphicx}
\usepackage{dcolumn}
\usepackage{bm}
\usepackage{color}
\usepackage[utf8]{inputenc}
\usepackage[T1]{fontenc}
\usepackage{mathptmx}
\usepackage{upgreek}

\begin{document}

\title[Anisotropic magnetotransport  in LaAlO$_3$/SrTiO$_3$ nanostructures]{Anisotropic magnetotransport  in LaAlO$_3$/SrTiO$_3$ nanostructures}

\author {Mithun S Prasad}
	\affiliation{Institut für Physik, Martin-Luther-Universität Halle-Wittenberg, Von-Danckelmann-Platz 3, 06120, Halle, Germany. }
\author{Georg Schmidt}
	\email{georg.schmidt@physik.uni-halle.de}
	\affiliation{Institut für Physik, Martin-Luther-Universität Halle-Wittenberg, Von-Danckelmann-Platz 3, 06120, Halle, Germany. }
	\affiliation{Interdisziplinäres Zentrum für Materialwissenschaften, Martin-Luther-Universität Halle-Wittenberg, HeinrichDamerow-Str.4, 06120, Halle, Germany}

\date{\today}

\begin{abstract}
A number of recent studies indicate that the charge conduction of the LaAlO$_3$/SrTiO$_3$ interface at low temperature is confined to filaments which are linked to structural domain walls in the SrTiO$_3$  with drastic consequences for example for the temperature dependence of local transport properties. We demonstrate that as a consequences of this current carrying filaments on the nano-scale the magnetotransport properties of the interface are highly anisotropic. Our magnetoresistance measurements reveal that the magnetoresistance in different nanostructures ($<500nm$) is random in magnitude and sign, respectively. Warming up nanostructures above the structural phase transition temperature (105K) results in the significant change in MR. Even a sign change of the magnetoresistance is possible. The results suggest that domain walls that are differently oriented with respect to the surface exhibit different respective magnetoresistance and the total magnetoresistance is a result of a random domain wall pattern formed during the structural phase transition in the SrTiO$_3$ at cool down.
\end{abstract}

\maketitle

\section{introduction}

Interfaces between complex oxides shows great potential for future electronics \cite{mannhart2010oxide}. Since the discovery of the high mobility electron gas at the interface of $ LaAlO_3$ (LAO) and $SrTiO_3$ (STO) in 2004 \cite{ohtomo2004high}, a lot of studies have been conducted. Those studies revealed more interesting properties of the interface such as two dimensional superconductivity, induced ferromagnetism, gate tunability , highly efficient spin-charge conversion etc. \cite{reyren2007superconducting,thiel2006tunable,brinkman2007magnetic,lesne2016highly}. Enhanced room temperature mobility of LAO/STO nanowires \cite{irvin2013anomalous} holds great promise towards its room temperature application. Studies conducted by Dubroka et al. \cite{dubroka2010dynamical} show that the confinement region of the electron gas extends into STO substrate.
A resistance anomaly at $\sim 80K$ and $\sim 160K$  has been previously reported in structures grown at low oxygen pressure. The origin of the anomaly is often linked to structural phase transitions in the STO \cite{meaney2019partial,schoofs2012impact,seri2013thermally,goble2017anisotropic} but also alternative explanations were discussed\cite{dubroka2010dynamical}. In 2017 Minhas et al. \cite{minhas2017temperature} showed that the resistance anomaly can also be observed in material grown at high oxygen pressure when a lateral confinement in a nanostructure exists.

In 2013 scanning SQUID microscopy studies \cite{kalisky2013locally} indicated that the interface exhibits channeled current flow at low temperatures. These current carrying channels are linked to structural domain walls in the STO substrate that appear below a structural phase transition from cubic to tetragonal at a temperature of 105 K  \cite{lytle1964x}. The distribution of these channels changes every time the interface is warmed above this transition temperature  \cite{lytle1964x}. Kalisky et al. \cite{kalisky2013locally} also observed that the redistribution of the channels is completely random. Further studies also confirm this presence of conducting domain walls \cite{honig2013local,frenkel2016anisotropic,ma2016local}. These conducting domain walls are aligned along the crystallographic directions [100], [010], [110] and [1$\bar{1}$0]. Furthermore Ma  et al. \cite{ma2016local} observed indications of insulating areas in the vicinity of the conducting domain walls.
In confined systems, however, the domain walls can start to massivle influence the transport properties. In 2017 Goble et al. demonstrated that the resistivity perpendicular to a domain wall is higher \cite{goble2017anisotropic} than along the domain wall.  Almost at the same time Minhas et al. \cite{minhas2017temperature} showed an even more drastic effect on transport in LAO/STO nanostructures. They observed a resistance peak in 100nm wide LAO/STO channels around the phase transition temperature when warming up the structure. This peak is explained by a filamentary conduction at low temperatures. In structures of nanoscale dimensions, the  number of charge carrying filaments is limited and can be as small as one. The area around the filaments, however, becomes more and more insulating at lower temperatures because all charge is collected at the respective domain wall(s). When during warm-up the phase transition occurs, the single conducting domain wall breaks and the formerly conducting channel becomes insulating. Only at higher temperatures the charge is redistributed and normal conductance is restored. These findings have lead to a new perspective on the physics of current flow at these interfaces.
It is important to realize that during the cool down process it is not possible that the current path is interrupted. If there is a domain wall in the current path the electric field will collect the charge there and create a conducting filament. If there is no domain wall, the charge distribution remains unchanged during cool down and the material retains more or less homogeneous conductivity. This is the main reason why the cool down curve does not show any sign of the domain walls. In large area samples the sheer number of domain walls makes the phase transition and the redistribution of charge undetectable. Only recently, more evidence for domain wall conduction was presented by \cite{krantz2020observation} who found a Hall-effect like transverse resistance at zero field that can be explained by asymmetrically distributed domain walls that lead to the appearance of transverse voltages upon current flow.

\begin{figure}[t]
 \includegraphics[width=.65\columnwidth]{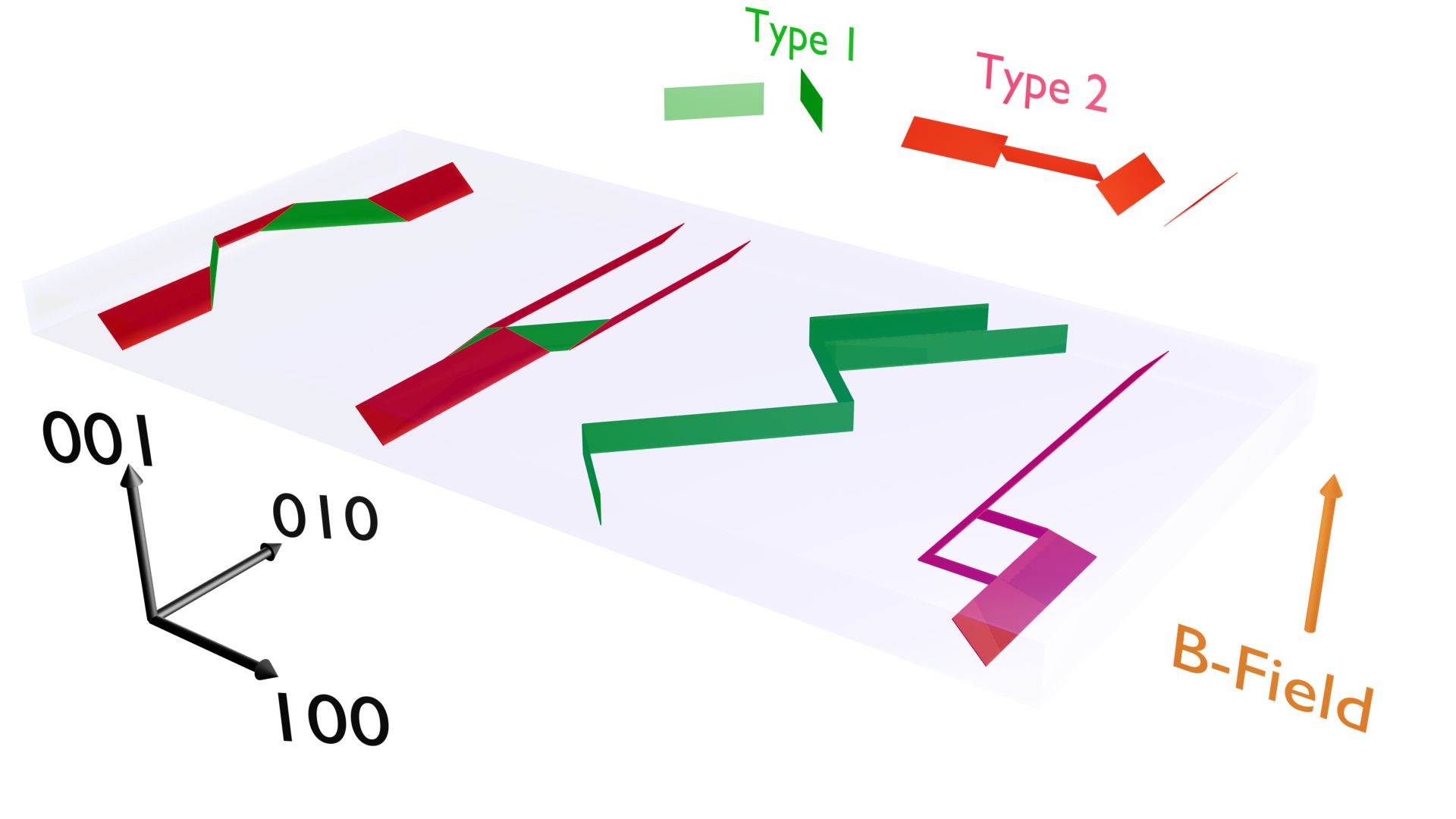}
  \caption{Alignment of type 1 and type 2 domain walls in the STO and possible random combinations. All four combinations connect the front and the back of the slab. Obviously it is possible to have combinations with different contributions from both types. Even for the same relative contribution different total lengths of the domain wall path can be realized. So even for same conduction direction we may have very different combinations of domain walls in the conducting channel that may also have different resistance depending on their respective lengths.}
  \label{fig:6}
\end{figure}

Taking this modified conductance mechanism into account several physical effects can be predicted that are expected for filamentary transport but that cannot be explained by the mainstream theory on LAO/STO interface conductivity. The presence or absence of these effects would give further insight into the transport physics in LAO/STO. Here the process to make high quality nanostructures that we have presented in \cite{minhas2016sidewall} opens up several options. To understand these experiments it is necessary to understand also the two different types of domain walls that can exist. One type (type 1) is oriented perpendicular to the surface and along the [110] and  [1$\bar{1}$0] crystalline directions in the subtrate plane. The other type is oriented at an angle  to the surface and along the [100]  and  [010] crystalline directions in the plane. As a consequence not only the resistivity of the two types of domain walls may be different but also a magnetic field perpendicular to the substrate surface that should be perpendicular to the 2DEG according to standard theory would be in plane of type 1 domain walls and tilted with respect to type 2 domain walls. This is crucial because in plane and (partly) perpendicular magnetic fields typically cause magnetoresistance that can be different in magnitude and/or sign, respectively. In addition it is important to note the respective in-plane orientation of the two domain wall types. With respect to the [100] direction Type 1 domain walls are oriented at 0$^\circ$ or 90$^\circ$ while type 2 domain walls are either oriented at 45$^\circ$ or 135$^\circ$. As Fig.\ref{fig:6} shows this allows to realize a domain wall pattern connecting both ends of a nanostructure that can be entirely composed by type 1 or type 2 domain walls alone or by a combination of both where any relative contribution of type 1 and type 2, respectively, is possible.

So if the conductivity in nanostructures is dominated by conducting filaments with random alignment we may expect the following effects.

1) Unless by accident the current path in a nanostructure will not consist of a single straight domain wall but of a chain of connected domain walls of different length, type and orientation. Thus the conduction path will have different respective lengths for each domain wall type. This must result in a random distribution of resistance values at low temperature for nominally identical nanostructures. Warming up through the phase transition and cooling down again should result in a different configuration and different resistance values for the same structure.
2) Because of the different types of domain walls and their possible alignment within the crystal lattice it may be possible that this randomness still shows some systematics with respect to the crystalline orientation of the nanostructures.
3) The magnetoresistance should also be different for the different structures. As shown above we can expect one type of domain wall which is perpendicular and another one which is tilted with respect to the surface. When we do magnetotransport measurements with the magnetic field applied perpendicular to the surface we would expect different response from the two types of domain walls. As mentioned, for perpendicular domain walls the magnetic field would be in-plane while for tilted domain walls it would also have a perpendicular component. As a consequence the MR response from different nominally identical nanostructures should differ and also change in a warm-up cool-down procedure.

We have designed a set of experiments in which a particular nanostructure design allows us to precisely verify/nullify these assumptions.

\begin{figure}[h!]
 \includegraphics[width=.65\columnwidth, height=9cm]{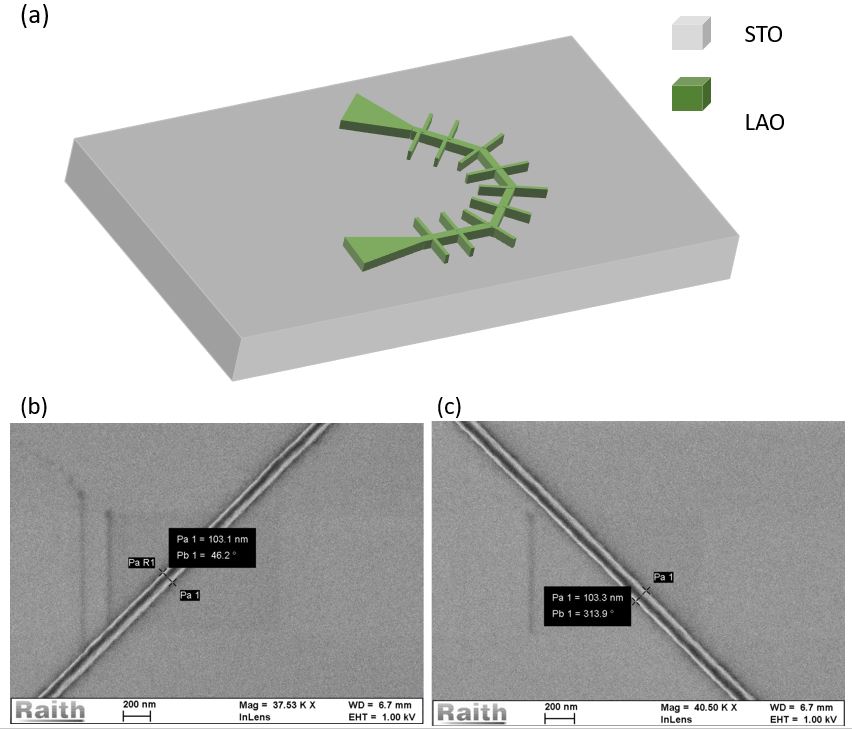}
  \caption
{(a) Sample geometry used for the measurements. The current-carrying path is divided into four equally long sections that are aligned at an angle of $ 0^0$, $45^0$, $ 90^0 $ and $135^0$ towards the substrate edge, respectively.(b),(c) Scanning electron microscope (SEM) image of two different orientation of 100nm structures ($45^0$ and  $135^0$ )}
\label{fig:1}
\end{figure}

\begin{figure*}[t!]
 \includegraphics[width=\linewidth]{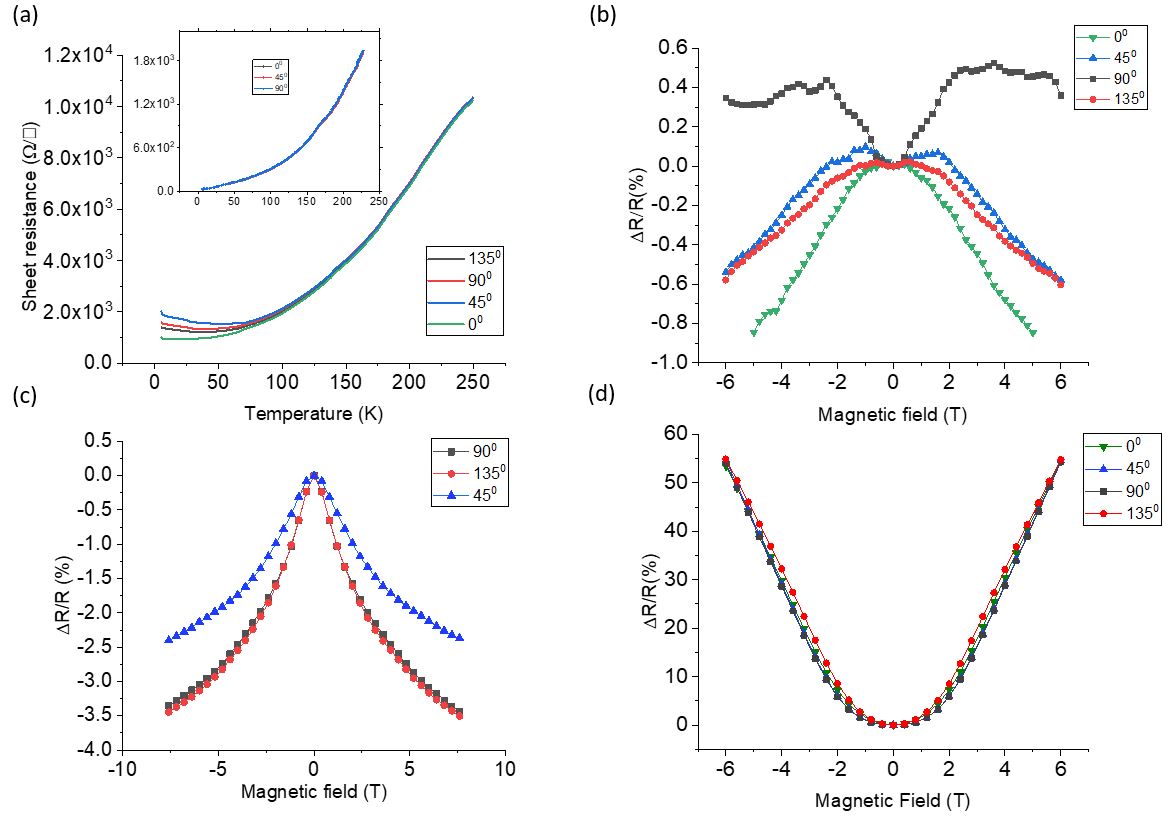}
\caption{(a) variation of resistance with temperature for different orientation of a 100nm structure (sample 1). A clear deviation of the resistance for differently aligned sections of the Hall bar appears when the temperature is below 100K. The insert shows the temperature dependence for a large area structure. (b-d) Anisotropy in magnetoresistance with orientation in 100nm structures from Sample1 (b), Sample 2 (c), and for a 2$\mu$m wide hall bar (d). For the nanostructures the MR is small compared to the 2$\mu$m structure, however the anisotropy that is predominant in the nanostructures almost vanishes in the larger Hall bar.}
  \label{fig:2}
 \end{figure*}

\section{fabrication}
The fabrication of the samples starts with the deposition of a 6 unit cell layer of LAO on $TiO_2$ terminated STO (001) using Pulsed laser deposition (PLD) in oxygen atmosphere with $p_{O_2}$ of $10^{-3}$ mbar at 850$^0$C. Laser fluence and pulse frequency are kept at 2 J/$m^2$ and 2 Hz, respectively. Reflection high energy electron diffraction (RHEED) is used to monitor the layer thickness with unit cell resolution during the growth. After deposition the sample is slowly cooled down to room temperature while the oxygen pressure is maintained. For nanopatterning we use the process originally published in \cite{minhas2016sidewall} which uses reactive ion etching (RIE) with $BCl_3$. With this process we are able to fabricate high quality nanostructures with lateral dimension down to 100 nm. The resulting patterned structures are stable at ambient conditions. The samples are bonded and electrical transport measurements are carried out in a $^4He$ bath cryostat with a variable temperature insert equipped with a superconducting magnet that allows a maximum magnetic field of 10T.
\begin{table}[ht]
\centering 											
\begin{tabular}{ | m{1.6cm} |  p{3.2cm} | p{3.2cm}|  } 								
\hline\hline 		\\[.3ex]									
Orientation   & Sheet resistance at 4.2K after initial cooldown    & Sheet resistance at 4.2K after warmup cycle \\ [1ex] 				

\hline 	\\											
 $ 0^0$ & 1 K$\Omega/\square$ & 2 K$\Omega/\square$  \\ 									
 $ 45^0$ & 2.1 K$\Omega/\square$ & 2.16 K$\Omega/\square$  \\
 $ 90^0$ & 1.6 K$\Omega/\square$ & 2.65 K$\Omega/\square$  \\
 $ 135^0$ & 1.4 K$\Omega/\square$ & 1.7 K$\Omega/\square$  \\ [1ex] 								
\hline 												
\end{tabular}
\caption{Shows the change in sheet resistance of 100nm structure after the warm up cycle to 200K (above the phase transition temperature 105K) }	
\label{table:resistance} 									
\end{table}

\begin{figure*}[t!]
 \includegraphics[width=\linewidth]{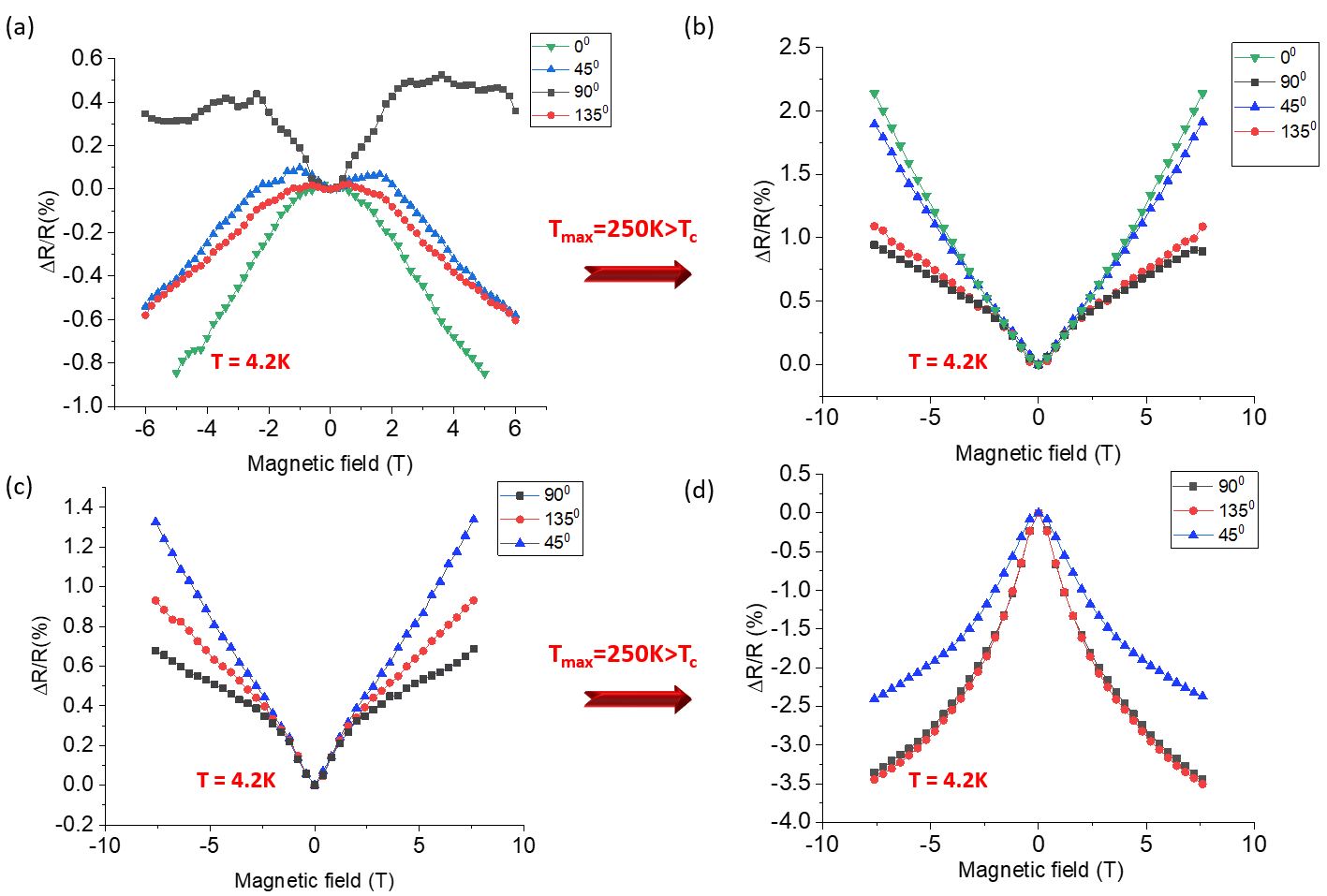}
\caption{Variation in MR for two different samples when warmup-cooldown cycle is performed to 250K. Both samples are heated to 250K and then wait for 1 hour and finally cooled down to 4.2K and then measurement was done at 4.2K.Both samples, sample 1 ((a) and (b)) and sample 2 ((c) and (d)) shows signifant change in MR behavior after the warmup through transition temperature.not only the magnitude but also the sign of MR changes.}
\label{fig:3}
\end{figure*}

\par

We used the sample geometry as shown in Fig. \ref{fig:1}a for the experiments. The structure consists of a nanopatterned Hall-bar of special design\cite{hupfauer2015emergence}. We have a continuous Hall-bar that consists of four connected segments, each with six voltage probes for measurement of longitudinal or transversal resistance. The segments are aligned at 0$^{\circ}$, 45$^{\circ}$, 90$^{\circ}$, and 135$^{\circ}$ with respect to the sample edge which corresponds to a major crystalline axis (100 or 010). The samples are cooled down at a rate of approx. 5 K/min and warm-up is done at a rate of approx. 2.5 K/min whenever required. The resistance is always measured in a four probe geometry. Voltages are measured using custom made zero drift voltage amplifiers and an Agilent 34420A 7.5 Digit nanovoltmeter. Current is measured by measuring the voltage over a series resistor of 1M$\Omega$. We apply a DC voltage of 100mV across the sample and the series resistor. Because of the design, we are able to measure the resistance of 3 nanostructures oriented at different respective angles simultaneously keeping all other parameters constant thus providing higher reproducibility and better comparability of the results. Also for better understanding and comparison we have included results from 2 nominally identical samples labeled sample 1 and sample 2.

\begin{figure*}[t!]
 \includegraphics[width=\linewidth]{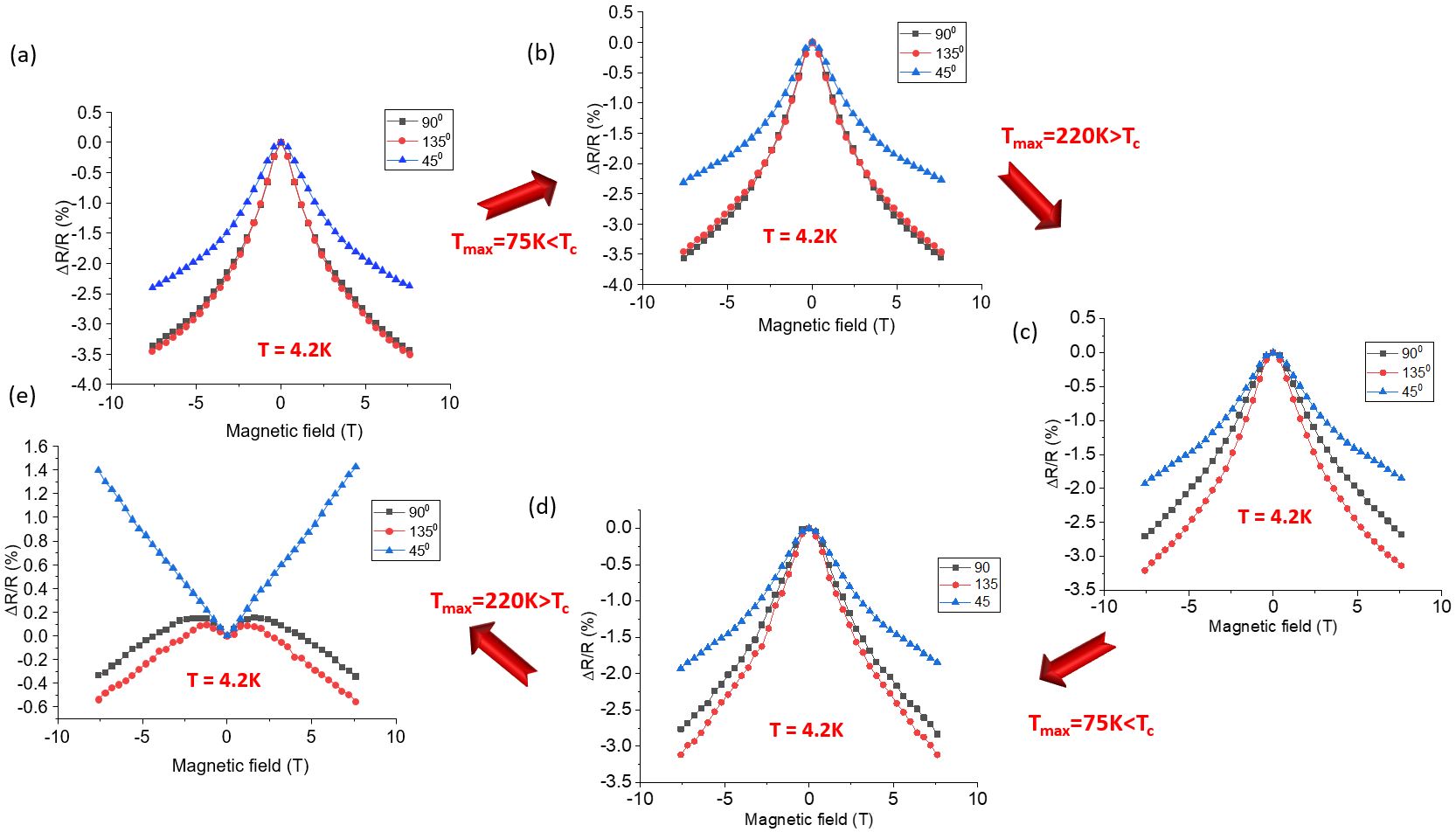}
\caption{\label{fig:4}Shows how the structural phase transition of STO (105K) influences the MR for sample 2. (a) shows the MR at 4.2K after initial cooldown. (b)shows MR behavior at 4.2K after warmup-cooldown cycle is performed to 75K< $T_c$ and (c) shows the MR at 4.2K after warmup-cooldown cycle to 220K > $T_c$.(d) No change in MR is observed subsequent warmup-cooldown cycle is performed to 75K< $T_c$ but (e) significant change in MR is observed when the sample is heated to 220K > $T_c$ above the phase transition.}
 \end{figure*}

\section{measurement}
As a first test we investigate the temperature dependence of the resistance for four nanostripes of different respective crystalline orientations using the hall bar geometry from Fig.\ref{fig:1}a for carrying out the resistance measurements.  Fig.\ref{fig:2}a shows the temperature dependence ( results from sample 1).

At room temperature there is no significant difference in resistance between the four different orientations of the nanostructures. Also during cool down the temperature decreases monotonically and identically for all four structures as expected but only down to the approximate temperature of the first structural phase transition of STO (105K). Below this temperature the four respective resistance values start to deviate. For three of the crystalline directions the resistance also starts to increase again below approx. 40 K, an effect that is not unknown also for large area structures. Cooling curves for a set of differently oriented structures of 2$\mu$m width do not show any anisotropy.
After this first confirmation we test whether warming and cooling again through the phase transition changes the result of the experiment. For sake of simplicity we only measure the resistance at 4 K. As table \ref{table:resistance} shows the resistance values for each individual stripe change significantly while they still remain different for all four stripes. This is consistent with a random formation of domain walls at the phase transition temperature.
We have repeated the experiment several times and also for different structures. Although we might have expected a preference for higher or lower resistance values for certain crystalline orientations, we cannot observe any significant preference for higher or lower values for any crystalline direction. This can be due to too low a number of experiments for valid statistics or due to the absence or smallness of the effect.

\par
In a next step we investigate the magnetoresistance (MR) in differently oriented nanostructures. It is well known that in a 2D or quasi 2D system like the domain walls the orientation of the magnetic field with respect to the plane is crucial for the magnetoresistance. For perpendicular magnetic fields we may simply have the so called geometric or Lorentz magnetoresistance which is positive and quadratic in magnetic field. This effect results from the Lorentz force that leads to (partially) circular electron orbits. For high fields and truly two dimensional systems one might even expect Shubnikov de Haas oscillations. For in plane fields, however, one would in first order expect no magnetoresistance. Nevertheless, it should be noted that for current flow in domain walls in $BiFeO_3$ a negative magnetoresistance has been observed for in plane fields\cite{he2012magnetotransport}. Because the type 1 and type 2 domain walls, respectively, are either perpendicular to the surface or tilted with respect to the surface, a field applied perpendicular to the surface is in-plane for the perpendicular domain walls and tilted with respect to the tilted ones. Similar to the statistical distribution of the resistance values in the first experiment we would thus expect also a statistical distribution of the magnetoresistance not only in magnitude but also in sign.

For the MR measurements the magnetic field is applied perpendicular to the sample surface. The field is swept from B=-6\,T to B=+6\,T. We first discuss the magnetoresistance for structures of 2$\mu$m width. Independent from the orientation, these structures show a large positive magnetoresistance which is quadratic in magnetic field as shown in Fig.\ref{fig:2}d. This signature is reminiscent of the Lorentz magnetoresistance discussed above. It should be noted that there are small differences in the MR for differently oriented stripes, nevertheless, the main contribution is identical for all orientations. For nanostructures we get a completely different picture. Fig.\ref{fig:2}b and Fig.\ref{fig:2}c shows two sets of MR curves taken on two different samples. The curves in each respective diagram were taken simultaneously for different parts of a single Hall bar with different respective crystalline orientations.
We immediately notice that the relative MR of the structures is much smaller than for the larger structure. The large Lorentz-type MR observed in the 2$\mu$m wide bars that showed a relative MR of $\Delta R/R>50\%$ at B=6\,T( Fig.\ref{fig:2}d) has almost completely vanished. The MR in the nanostructures is smaller than 1\% for the first structure and a few \% for the other one. Furthermore sign and/or magnitude of the MR differ for all orientations. The similarity of the curves for 45 and 135 degree for the first structure is purely random and cannot be reproduced on other samples. The fact that the magnitude of the MR in a large area structure  Fig.\ref{fig:2}d is at least one order of magnitude higher than for the nanostructures indicates that the physics associated with the origin of MR can be different.

\begin{figure*}[t!]
 \includegraphics[width=\linewidth]{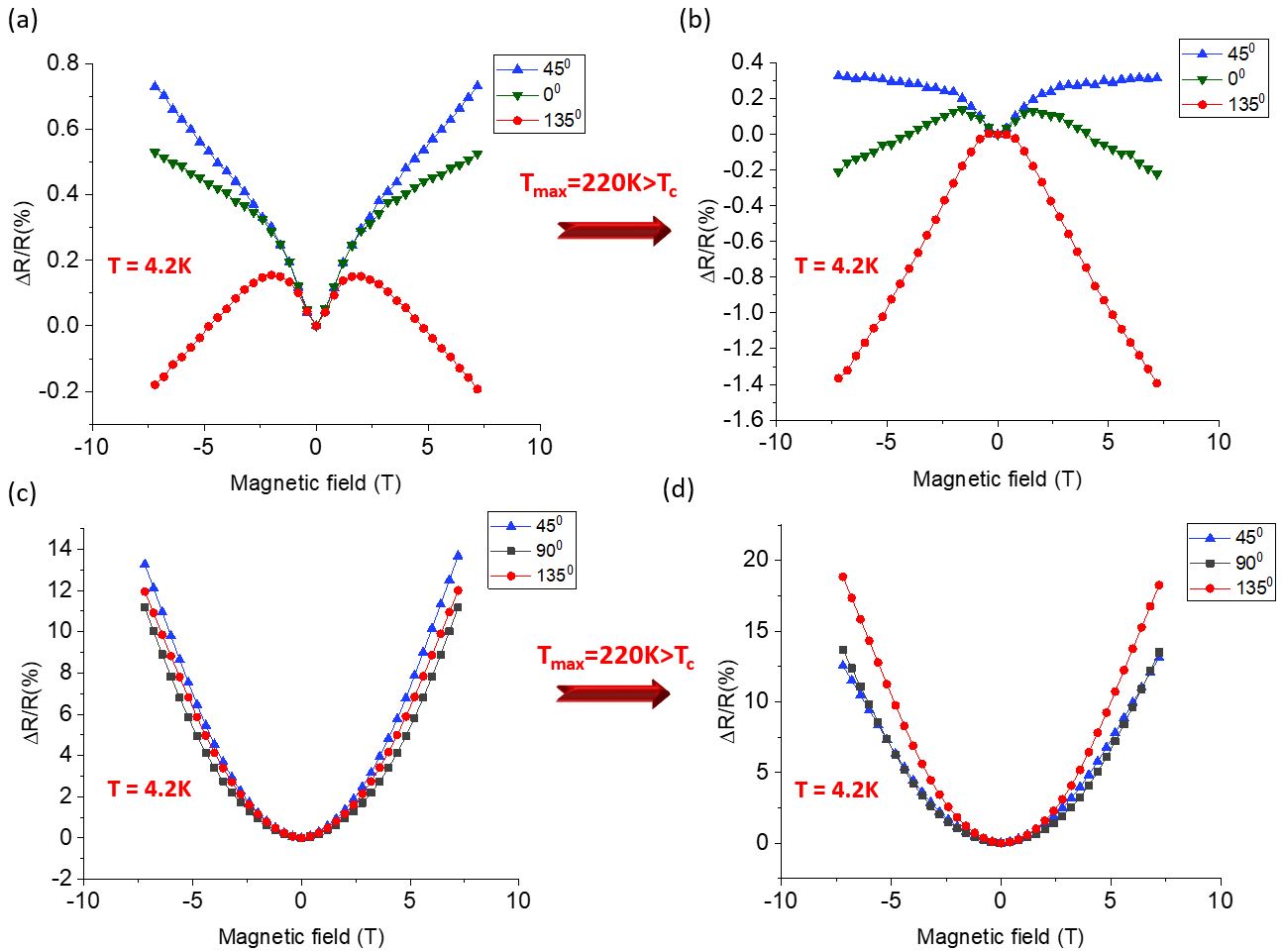}
  \caption{Shows how the MR measured at 4.2K for (a) 200nm structure and (c) 300nm structure. 300nm structure behaves more similar to large area structure but 200nm structures resembles more with 100nm structure. Heating through the transition temperature results in change in MR for both 200nm (b) and 300nm (d) structure.}
  \label{fig:5}
\end{figure*}

In a next step we investigate the MR after warm up and cool-down through the temperature of the structural phase transition. Both structures were warmed up to 250K Fig.\ref{fig:3} shows the results for structures 1 and 2, respectively after warming up to 250 K and cooling down again to 4.2 K. In both cases the MR has changed considerably. Especially for structure 2 we observe a change in magnitude and sign for all three measured directions. Further temperature sweeps show that the phase transition seems to be crucial for the effect. Fig.\ref{fig:4} shows a further sequence of MR measurements after different warm-up and cooling cycles for structure 2. After the measurement from Fig.\ref{fig:3} the sample was first warmed up to 75 K which is below the phase transition temperature. The sample was kept at this temperature for 60 min. and then cooled down to 4.2 K again. The resulting MR measurements look identical (Fig.\ref{fig:4}b) to the previous ones. The sample is then warmed up to 220 K which is above the transition temperature. After again cooling down to 4.2 K we observe a moderate change in the magnitude of the MR, especially for an angle of 90 degree (Fig.\ref{fig:4}c). Repeating the same sequence warming up to 75 K for one hour and cooling down yields no change (Fig.\ref{fig:4}d). Warming up again through the phase transition (T=220K) for one hour, however, results in a massive change (Fig.\ref{fig:4}e). For 90 and 135 degree the MR is reduced by a factor of $\approx$7. For the 45 degree direction, however, we observe a complete sign reversal of the MR.

As pointed out in Minhas et al.\cite{minhas2017temperature} the effect of the domain walls is very pronounced for structures as small as 100 nm but it averages out when the lateral size of the structures is even moderately increased to a few hundred nm. The investigation of further samples with lateral dimensions of 200 and 300 nm, respectively, confirms this statement.

In  Fig.\ref{fig:5} we show the MR for 200nm and 300nm wide nanostructures and the change in MR when heating through the transition temperature. For the 200nm structure ( Fig.\ref{fig:5}a and  Fig.\ref{fig:5}b), the MR obtained was hugely direction dependent and has lower magnitude just like in the case of the 100nm structure. After cycling the temperature through T$_{PT}$ a significant change is observed. For a 300nm wide structure( Fig.\ref{fig:5}c and  Fig.\ref{fig:5}d), however, the MR is larger in magnitude and quadratic in nature, as was the case for the $2\mu$m structure ( Fig.\ref{fig:2}d). After sweeping the sample temperature through T$_{PT}$, the main quadratic behavior remains unchanged and only the small changes in magnitude appear as we might expect for a sample that is still close to the critical size.

\section{discussion}

We now discuss the different predictions that we made in the beginning:
1) We certainly observe the random scattering of resistance values not only for different crystalline directions. Even in a single direction cycling the temperature through the phase transition leads to a change in low temperature resistance. The different possible domain wall configurations may also be the cause for the non monotonicity of the resistance/temperature curve that sometimes is observed and sometimes not. It is easily understood that a rearrangement of the carrier distribution can lead to an increase in resistance depending on the domain wall properties.
2) At least with the number of experiments that were carried out we cannot see a correlation between resistance and crystalline direction. This is most likely due to the randomness of the domain wall distribution.
3)Up to a width of 200 nm we observe also a random distribution of possible MR curves within a certain range. In a single structure the MR curve can vary between different crystalline orientations. Nevertheless, again we observe no correlation of one direction with special MR characteristics. as in 1) the effect also changes when the temperature is cycled through T$_{PT}$ but it does not change for a temperature cycle that remains below T$_{PT}$. The MR that we observe in nanostructures can be positive or negative which is consistent with two different types of domain walls with two different orientations, namely perpendicular or tilted, that can lead to positive or negative magnetoresistance, respectively. As shown in  Fig.\ref{fig:6} the crystalline alignement of the two types of domain walls allows for a large variation of the respective contributions to the resistance and magnetoresistance. These different relative contributions explain well the observed behavior.
Also the transition to a Lorentz type of magnetoresistance in larger structures is readily explained by our model. In a small structure the conductin path merely consists of a chain of domain walls that does not allow for circular electron orbits due to the magnetic field. In a larger structure we may have a two dimensional network of domain walls where the mesh size may be suitable for circular electron motion that soon dominates the magnetoresistance.
Unfortunately it is not possible to design a simple model that describes quantitatively the resistance values based on two types of domain walls with fixed resistivity. Among other aspects the conductivity of the domain walls is determined by the respective number of carriers in the domain wall. Assuming that the initial number of carriers is constant, different domain wall configurations with different total length and different contributions from type 1 and type 2 must have different conductivity and will thus contribute differently to the MR. It is even unclear whether the carrier concentration is homogeneous through the domain walls or can vary between different positions.

\section{conclusion}

As a conclusion we may state that in our experiments we find effects that can be predicted based on the theory of filamentary transport but that do not fit the model of a quasi 2D electron gas with sheet conductivity at the LAO/STO interface. Both resistance and magnetoresistance have random values within a certain range consistent with the formation of two different types of conducting domain walls with different orientation with respect to the surface. The results raise the question whether describing the transport properties of large area structures is possible using a simple  band structure or whether the microscopic domain wall structure needs to be taken into account.

\section{acknowledgement}
We wish to acknowledge the support of the DFG in SFB762, project B09

\end{document}